\documentclass[twocolumn,times]{aastex6}

\usepackage{natbib}
\usepackage{placeins}
\usepackage{amsmath}
\usepackage{mathtools}
\newcommand\degree{{^\circ}}
\newcommand\kms{km~s$^{-1}$}
\defcitealias{vogt1995}{V95}
\defcitealias{koch2007b}{KK07}
\defcitealias{kirby2010}{KG10}
\defcitealias{spencer2017a}{Paper I}

\begin{document}

\title{The Binary Fraction of Stars in Dwarf Galaxies: the Case of Leo II}
\shorttitle{Binary Fraction of Leo II}
\shortauthors{Spencer et al.}
\author{Meghin E. Spencer\altaffilmark{1}, Mario Mateo\altaffilmark{1}, 
		Matthew G. Walker\altaffilmark{2}, Edward W. Olszewski\altaffilmark{3}, 
		Alan W. McConnachie\altaffilmark{4}, \\ Evan N. Kirby\altaffilmark{5}, and 
		Andreas Koch\altaffilmark{6}}
\altaffiltext{1}{Department of Astronomy, University of Michigan, Ann Arbor, MI, USA, meghins@umich.edu}
\altaffiltext{2}{McWilliams Center for Cosmology, Department of Physics, Carnegie Mellon University, Pittsburgh, PA, USA}
\altaffiltext{3}{Steward Observatory, University of Arizona, Tucson, AZ, USA}
\altaffiltext{4}{NRC Herzberg Institute of Astrophysics, Victoria, BC, Canada}
\altaffiltext{5}{Department of Astronomy, California Institute of Technology, Pasadena, CA, USA}
\altaffiltext{6}{Department of Physics, Lancaster University, Lancaster, UK}

\begin{abstract}
We combine precision radial velocity data from four different published works of the stars in the Leo II dwarf spheroidal galaxy. This yields a dataset that spans 19 years, has 14 different epochs of observation, and contains 372 unique red giant branch stars, 196 of which have repeat observations. Using this multi-epoch dataset, we constrain the binary fraction for Leo II. We generate a suite of Monte Carlo simulations that test different binary fractions using Bayesian analysis and determine that the binary fraction for Leo II ranges from  $0.30^{+0.09}_{-0.10}$ to $0.34^{+0.11}_{-0.11}$, depending on the distributions of binary orbital parameters assumed. This value is smaller than what has been found for the solar neighborhood ($\sim$0.4--0.6) but falls within the wide range of values that have been inferred for other dwarf spheroidals (0.14--0.69). The distribution of orbital periods has the greatest impact on the binary fraction results. If the fraction we find in Leo II is present in low-mass ultra-faints, it can artificially inflate the velocity dispersion of those systems and cause them to appear more dark matter rich than in actuality. For a galaxy with an intrinsic dispersion of 1 \kms{} and an observational sample of 100 stars, the dispersion can be increased by a factor of 1.5-2 for Leo II-like binary fractions or by a factor of 3 for binary fractions on the higher end of what has been seen in other dwarf spheroidals.
\end{abstract}


\section{Introduction}

The first measured velocity dispersion for a dwarf galaxy was found to be 6.5 \kms{} based on only four stars in Draco \citep{aaronson1983}. This corresponded to a mass to light ratio of 31, which indicated that Draco was embedded in a dark matter halo. Because this was such a radically different ratio than that of globular clusters devoid of dark matter, five common theories involving galactic tides, small number statistics, poor velocity precision, stellar atmospheric jitter, or binary stars were proposed to potentially explain the large velocity dispersions without the need for dark matter \citep{aaronson1983,cohen1983,mcclure1984}.

One explanation was that Draco was being tidally disrupted by the Milky Way, as velocity dispersion is only a good mass estimator if the galaxy is in dynamic equilibrium. This seemed plausible since Draco is the closest of the classical dSph galaxies to the Milky Way. However, with the addition of radial velocity data from other dSphs---Sculptor \citep{armandroff1986}, Ursa Minor \citep{aaronson1987,armandroff1995,olszewski1995}, Fornax \citep{mateo1991}, Carina \citep{mateo1993}, Sextans \citep{suntzeff1993,hargreaves1994a}, Leo II \citep{vogt1995}, and Leo I \citep{mateo1998}---it became apparent that most dwarfs, regardless of their proximity to the Milky Way, exhibited large velocity dispersions without evidence for streaming motions. In addition, some simulations predicted that a perigalactic passage would leave behind a velocity gradient larger than the velocity dispersion \citep{piatek1995,pryor1996}, a feature that is not seen in any of the aforementioned dwarfs. This initially seemed to rule out the explanation of tides, but other simulations have shown that the fast stellar kinematics of dSphs might be produced through repeated tidal shaping of a more massive progenitor by the Milky Way \citep{kroupa1997,klessen1998}. The remnants of these interactions do not always exhibit tidal tails and when observed at the right time along the right orbit they can produce dwarf galaxies equivalent to what is observed \citep{casas2012}.

While completely ruling out tides has been difficult, the growth of the spectroscopic surveys did eliminate the concern over small number statistics as the number of stars per dSph increased from four to several hundred. Furthermore, state of the art spectrographs can now measure velocities at 1--2 \kms{} precision, making it possible to extract the dispersions in ultra-faints, which are only a few \kms{}. The advent of better spectrographs also allowed for the observation of fainter K-giants, which exhibit far less atmospheric jitter than brighter carbon stars \citep{mayor1984,seitzer1985}.

The fifth and final theory was that radial velocity components from binary stars were contributing to the velocity dispersion. Repeat observations of Draco stars showed that binaries contributed very little to the high velocity dispersion \citep{aaronson1987,olszewski1995}, and Monte Carlo simulations of binaries predicted the same results \citep{hargreaves1996,olszewski1996}. Furthermore, studies of Ursa Minor \citep{olszewski1995}, Sculptor \citep{queloz1995}, and Leo II \citep{koch2007b} saw indistinguishable changes in dispersions measured from one epoch of velocity data versus multiple epochs. All in all, the addition of more and better velocity measures has mitigated most of the skepticism surrounding these large velocity dispersions. As such, it is now widely accepted that dSphs are some of the most dark matter dominated objects in the universe.

However, when it comes to binary stars, the question of contamination is still somewhat open ended for the more recently discovered ultra-faint dwarfs. These faint galaxies have dispersions closer to the 2--3 \kms{} that can be contributed by binaries, making them more susceptible to velocity dispersion inflation. Recent work by \citet{dabringhausen2016} has verified that binaries affect the inferred properties of ultra-faints to a greater extent than their more massive counterparts. It was also shown by \cite{mcconnachie2010} that there is a $\ga20\%$ chance that the intrinsic velocity dispersions of many ultra-faints (e.g., Segue 1, Segue 2, Willman 1, Bootes II, Leo IV, Leo V, and Hercules) are actually $\sim$0.2 \kms{} like globular clusters, but the presence of binaries has increased the observed dispersions to a few \kms{}. While this is an extreme scenario, the fact that binary stars can drastically impact the velocity dispersion of ultra-faints cannot be ignored. 

For the galaxy Bootes I, \cite{koposov2011} repeatedly took spectra of the same stars 15 times over the course of one month and discarded any stars that showed velocity variability. As a result, they found that the stars in Bootes I could be fit by a single population having a velocity dispersion of $4.6^{+0.8}_{-0.6}$ \kms{}, as opposed to previous single-epoch velocity dispersion measurements of $6.6\pm2.3$ \kms{} \citep{munoz2006b} and $6.5^{+2.0}_{-1.4}$ \kms{} \citep{martin2007}. While this is a significant step in the right direction, simply removing the velocity variables does not remove all the binaries, as there can be stars with orbital periods much longer than the observation cadence. In Segue 1, \cite{simon2011} not only removed obvious velocity variables to get a dispersion of $3.9\pm0.8$ \kms{}, but they also corrected for binaries that were non-variable on the timescale of their observations, finding a slightly lower dispersion of $3.7^{+1.4}_{-1.1}$ \kms{} \citep{simon2011,martinez2011}. For comparison, the single-epoch velocity dispersion was measured at $4.3\pm1.2$ \kms{} \citep{geha2009}.

\cite{minor2010} describe a generalized method to estimate the binary fraction and remove the effects of long-period binaries on the velocity dispersion for any dwarf galaxy, as was done in \cite{simon2011} for Segue 1. One major pitfall of this is the need for multi-epoch observations, which are currently not available for most ultra-faints. To make matters worse, this analysis also necessitates velocity errors of $\le1$ \kms{} due to the already small velocity dispersions \citep{mcconnachie2010,minor2010}. There is no doubt that such observations will become available in the future, but an alternative approach that can be used in the interim is to provide a range of plausible intrinsic velocity dispersions for ultra-faints based on the binary fractions in classical dwarfs. In this way, we can predict how big of an effect binaries could have on ultra-faints. 

A detailed binary analysis has been performed on Carina, Fornax, Sculptor, and Sextans \citep{minor2013}, but not for Draco, Ursa Minor, Leo I, and Leo II. In this paper, we turn our attention to Leo II. Relatively few spectroscopic observations have been taken for this dwarf galaxy due to its large distance away from the Milky Way \citep[233$\pm$15 kpc,][]{bellazzini2005}. \cite{spencer2017a} significantly expanded upon preexisting data by adding radial velocities from MMT/Hectochelle for 175 member stars over the course of eight years with as many as five observational epochs per star. Combining this with other studies \citep{vogt1995,koch2007b,kirby2010} now makes it possible to perform an extensive analysis on the binary fraction in Leo II. 

This paper is organized as follows. In Section \ref{sec_data}, we introduce the dataset for Leo II. In Section \ref{sec_methods}, we describe the methodology for determining the binary fraction of a dwarf galaxy. Section \ref{sec_leoiiresults} contains the results for Leo II and Section \ref{sec_ultrafaints} quantifies the implications for ultra-faints.  The summary and conclusions are in Section \ref{sec_conclusion}.

\section{Radial Velocities}\label{sec_data}

We use radial velocity data from four studies, which are summarized in Table \ref{table_datasets}. The first set comprises 31 red giant branch (RGB) stars with a median radial velocity error of 3 \kms{} \citep[][hereafter V95]{vogt1995}. It contains the first spectroscopic observations of RGB stars in Leo II, and remained the only kinematic dataset for over a decade. The second study, by \citet[][hereafter KK07]{koch2007b}, consists of radial velocities for 171 member stars. \citetalias{koch2007b} published average velocities taken during three epochs between 2003 and 2004. Velocity measures that are averaged over more than a few days (as in \citetalias{koch2007b}) will damp out the velocity changes caused by binaries. Instead, we used the unpublished single-epoch velocity measures, which were taken on the three dates listed in \citetalias{koch2007b}. We have included these velocities in Table \ref{data_table}. The drawback of using the non-averaged velocities in \citetalias{koch2007b} is that the error bars can be very large (up to $\sim$140 \kms{}). We chose to exclude \citetalias{koch2007b} measurements with errors larger than 35 \kms{} or $\chi^{2}$ from the average larger than 3. This removed 20 measurements from the two epochs in 2003 and leaves us with a median velocity error of 2.8 \kms{}.

The third dataset comes from \citet[][hereafter, KG10]{kirby2010}. They used Keck/DEIMOS to obtain medium resolution spectroscopy for the purpose of chemical abundance measurements, but also extracted radial velocities to help identify member stars. This was done by cross-correlating the red half of each spectrum with a set of template spectra from \citet{simon2007}. The cross-correlation peak from the best fitting spectrum was adopted as the velocity. Velocity errors were calculated by resampling the spectrum 1000 times with different noise realizations. The error was the quadrature sum of the systematic error floor \citep[2.2 \kms{},][]{simon2007} and the standard deviation of the 1000 velocity trials. These measurements were not published in \citetalias{kirby2010}, so we include them in Table \ref{data_table}. Additional details of the observations can be found in \citetalias{kirby2010}. This dataset contains one epoch of velocities for 258 stars with a median error of 2.3 \kms{}.

\begin{figure}
\epsscale{1.1}
\plotone{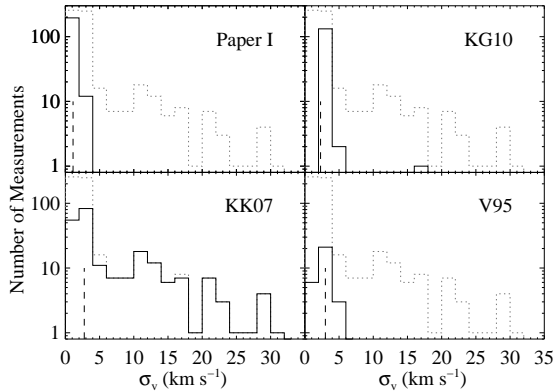}
\caption{Histograms of the radial velocity errors for each of the four datasets are shown as solid black lines. Vertical black dashed lines show the median error for each of the four datasets. The gray dotted line is the histogram of the velocity errors for the combined data set. Only measurements for stars with more than one observation are plotted (i.e., the measurements in Table \ref{data_table}).}
\label{errors}
\end{figure}

The fourth and final dataset is published in \citet[][hereafter, Paper I]{spencer2017a}, which contains radial velocities for 175 member stars. 50 of these have two or more observations, which were taken over the course of 8 years with Hectochelle \citep{szentgyorgyi1998} on the Multiple Mirror Telescope. This dataset contains 5 epochs between the years 2006 and 2013. The median error for these velocities is 1.1 \kms{}. Histograms of the error bars for each of these four studies are shown in Figure \ref{errors}. 

We note that \cite{bosler2007} reported velocities for 74 stars, but since their focus was on stellar chemistry rather than kinematics and their radial velocity errors are $\sim$50 \kms{}, the data are not precise enough for us to use in this study.

\begin{figure}
\epsscale{1.1}
\plotone{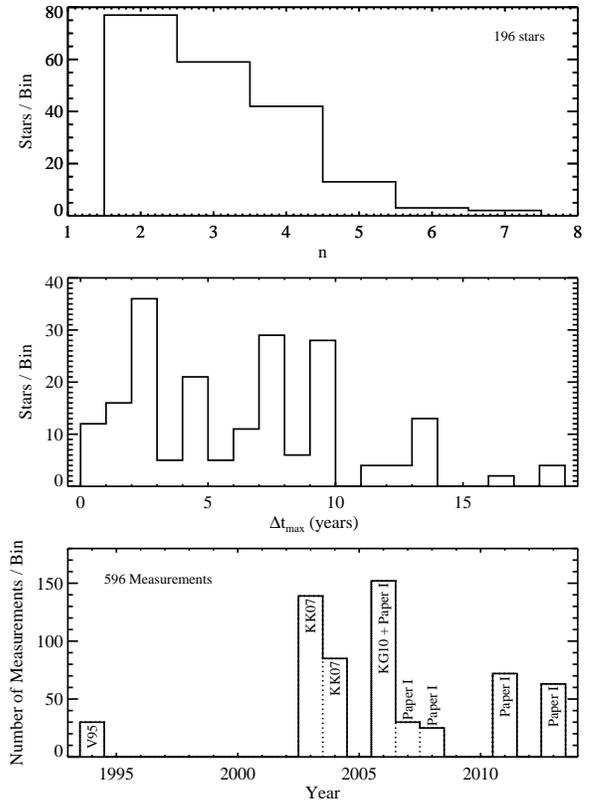}
\caption{Top: number of observations per star for the 196 stars in the sample. Middle: longest time separation between measurements per star. Bottom: number of velocity measurements per year for the 596 measurements in the sample. Bins are labeled with the paper that contributed those measurements.}
\label{nobs_deltat_year}
\end{figure}

Taking these four datasets together, the total number of unique RGB stars with multiple observations in Leo II is 196. In Figure \ref{nobs_deltat_year}, we plot some useful quantities to help summarize this larger dataset. The top panel shows the number of observations per star, with the maximum being 7 observations. The middle panel has the maximum time baseline for each star. This ranges from 11 days to nearly 19 years. Finally the bottom panel shows the number of observations taken per year. The years are labeled with the study that contributed to them. In total, we have 596 independent velocity measurements. Table \ref{table_datasets} summarizes the systemic velocity, velocity dispersion, median velocity error, number of stars, and number of epochs contained in each of the four studies.

Table \ref{data_table} lists the measurements that we used in this study. Column 1 is an id number that we assign to each unique star. Column 2 is the number of observations for that star. Columns 3 and 4 contain the coordinates. Column 5 lists the Heliocentric Julian date when the observations were made. Column 6 has the radial velocity and uncertainty after adjusting for any systematic offsets (see the next paragraph). Column 7 lists the relevant paper. Measurements from \citetalias{vogt1995} and \citetalias{spencer2017a} have been previously published, whereas measurements referencing \citetalias{koch2007b} and \citetalias{kirby2010} have not. Only stars that had more than one observation are included in the table.

\begin{figure}
\epsscale{1.1}
\plotone{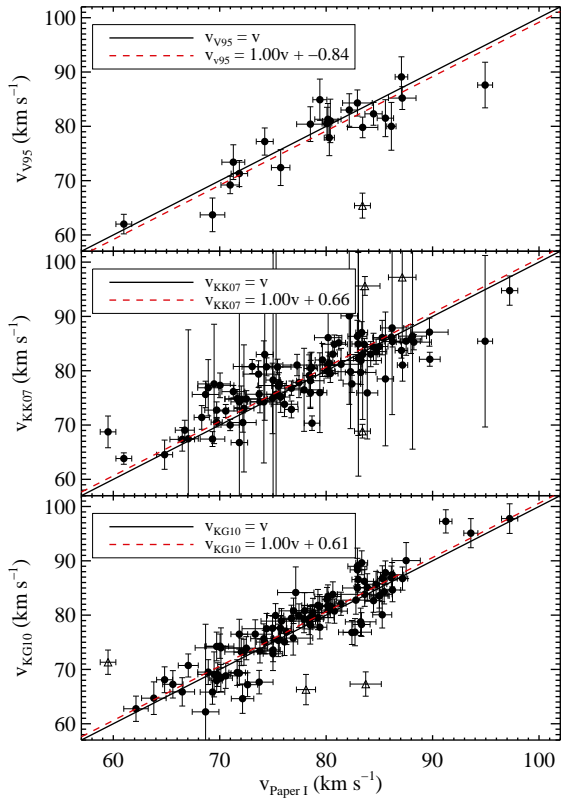}
\caption{Radial velocities measured by this paper versus velocities measured by other papers. Top: \citetalias{vogt1995} had 23 stars that overlapped with our survey. Middle: \citetalias{koch2007b} had 97 overlapping stars. Bottom: \citetalias{kirby2010} had 97 overlapping stars. The solid black line indicates where stars would be if the measurements perfectly matched. The dashed red line indicates the best fit to the data after setting the slope equal to 1. Stars that had different velocities by more than 10 \kms{} were not included in the fit, and are shown by open triangles. The \textit{y}-intercept of this line is the systematic offset between the datasets and was subtracted from the respective datasets.}
\label{compare_vels}
\end{figure}

As a consequence of combining data from different spectroscopic surveys, we needed to identify if there were any systematic offsets present between the studies. Figure \ref{compare_vels} shows average velocities from \citetalias{spencer2017a} plotted against the average velocities reported in \citetalias{vogt1995}, \citetalias{koch2007b}, and \citetalias{kirby2010} when stars existed in both catalogs. For each comparison, we fit a line weighted on the ordinate errors and set the slope equal to 1. Stars with velocities that disagreed by more than 10 \kms{} were excluded from the fit. Such stars pulled the fit lines away from the main group of stars, especially since they all had small error bars, as was found by inspection. The seven stars that fall into this category are plotted as open triangles. Finally, we took the resulting \textit{y}-intercept of the best-fit line as the systematic offset between the external datasets and our dataset in \citetalias{spencer2017a}. We subtracted these corrections, such that the corresponding velocities follow the form $v_{study\_corrected} = v_{study} -$\textit{offset}. The offset values are -0.84 \kms{} for \citetalias{vogt1995}, 0.66 \kms{} for \citetalias{koch2007b}, and 0.61 \kms{} for \citetalias{kirby2010}. 

\begin{deluxetable*}{l c c c c}
\tablecaption{Summary of published velocity data\label{table_datasets}}
\tabletypesize{\scriptsize}
\tablehead{\colhead{Measure} & \colhead{\citetalias{spencer2017a}}  & \colhead{\citetalias{kirby2010}} & \colhead{\citetalias{koch2007b}} & \colhead{\citetalias{vogt1995}} }
\startdata
Systemic Velocity (\kms{}) & 78.3$\pm$0.6 & Not Reported & 79.1$\pm$0.6 & 76$\pm$1.3 \\
Velocity Dispersion (\kms{})& 7.4$\pm$0.4 & Not Reported & 6.6 $\pm$0.7 & 6.7$\pm$1.1 \\
Median Velocity Error (\kms{}) & 1.1 & 2.3 & 2.8 & 3.0 \\
Number of Stars & 175 & 258 & 171 & 31 \\
Number of Epochs & 5 & 1 & 3 & 1 \\
\enddata
\end{deluxetable*}

\begin{deluxetable*}{c c c c c r c}
\centering
\tablecaption{Velocities of RGB stars in Leo II\label{data_table}}
\tabletypesize{\scriptsize}
\tablehead{\colhead{Star ID} & \colhead{n} & \colhead{$\alpha_{\mathrm{J2000}}$} & \colhead{$\delta_{\mathrm{J2000}}$} & \colhead{HJD} & \colhead{$v$\tablenotemark{a}} & \colhead{Ref.}  \\
& & \colhead{(hh:mm:ss.ss)} & \colhead{(dd:mm:ss.ss)} & \colhead{(days)} & \colhead{[\kms]} & }
\startdata
LeoII-016 & 4 & 11:13:32.05 & 22:08:58.61 & 2453850.7 &   70.98$\pm$  0.86 &  Paper I \\
LeoII-016 & 4 & 11:13:32.07 & 22:08:58.70 & 2452693.2 &   71.80$\pm$  4.21 &     KK07 \\
LeoII-016 & 4 & 11:13:32.07 & 22:08:58.70 & 2453061.2 &   69.19$\pm$  1.07 &     KK07 \\
LeoII-016 & 4 & 11:13:32.01 & 22:08:58.40 & 2449431.9 &   70.04$\pm$  1.60 &      V95 \\
LeoII-017 & 5 & 11:13:27.69 & 22:10:39.76 & 2454212.8 &   80.39$\pm$  0.73 &  Paper I \\
LeoII-017 & 5 & 11:13:27.70 & 22:10:39.90 & 2452693.2 &   72.77$\pm$  7.36 &     KK07 \\
LeoII-017 & 5 & 11:13:27.70 & 22:10:39.90 & 2453061.2 &   79.13$\pm$  1.68 &     KK07 \\
LeoII-017 & 5 & 11:13:27.69 & 22:10:39.90 & 2453770.0 &   81.09$\pm$  2.14 &     KG10 \\
LeoII-017 & 5 & 11:13:27.67 & 22:10:39.27 & 2449432.1 &   81.84$\pm$  4.00 &      V95 \\
LeoII-018 & 4 & 11:13:29.46 & 22:09:49.46 & 2454212.8 &   85.57$\pm$  0.77 &  Paper I \\
LeoII-018 & 4 & 11:13:29.47 & 22:09:49.61 & 2452693.2 &   77.83$\pm$ 12.32 &     KK07 \\
LeoII-018 & 4 & 11:13:29.46 & 22:09:49.60 & 2453770.0 &   87.22$\pm$  2.15 &     KG10 \\
LeoII-018 & 4 & 11:13:29.43 & 22:09:49.04 & 2449432.0 &   82.34$\pm$  3.40 &      V95 \\
\enddata
\tablenotetext{a}{Velocities after correcting for systematic offsets. Only stars with multi-epoch velocity measurements are included.}
\tablecomments{This table is published in its entirety in the electronic edition of the Astronomical Journal. A portion is shown here for guidance regarding its form and content.}
\end{deluxetable*}

\subsection{Velocity variability}

Although the goal of this paper is to determine the binary fraction of the galaxy, we can also use our dataset to single out individual stars that are binary candidates. These stars will show velocity variability that cannot be accounted for by the velocity measurement uncertainties.

\begin{figure}
\epsscale{1.1}
\plotone{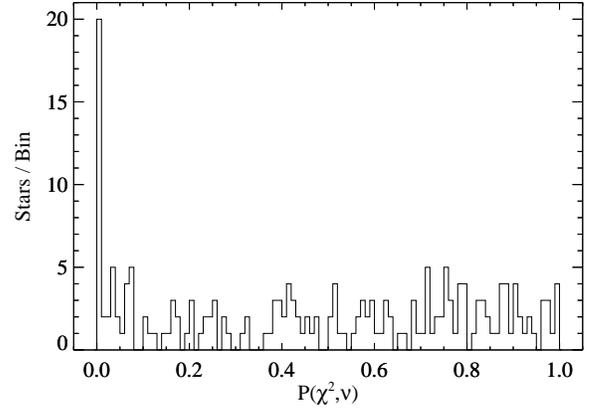}
\caption{Probability of exceeding $\chi^{2}_{\nu}$ for each star. Stars that are likely binaries will have $P(\chi^{2},\nu)<0.01$. 20 stars fall into this region and the expectation is only 2.}
\label{binary_chisqu}
\end{figure}

For each star with multiple observations, we calculated the reduced chi squared statistic as
\begin{equation}
\chi^{2}_{\nu} = \frac{1}{\nu} \sum_{i}^{n}\Big(\frac{v_{i}-\langle v\rangle}{\sigma_{i}}\Big)^{2} ,
\end{equation}
where $\nu = n-1$ is the number of degrees of freedom, and $n$ is the number of observations per star. For reference, the number of stars with a given $n$ is plotted in Figure \ref{nobs_deltat_year}. The probability of exceeding a given $\chi_{\nu}^{2}$ is $P(\chi^{2},\nu)$. A histogram of these probabilities is shown in Figure \ref{binary_chisqu}. If no binaries are present, then this distribution should be uniform over all probabilities, which equates to about 2 stars per bin. Alternatively, if binaries are present, they would cause a spike in the number of stars with low probability. The latter case is precisely what we see in Figure \ref{binary_chisqu}. The bin with $P(\chi^{2},\nu)<0.01$ contains 20 stars rather than the null hypothesis of 2 stars. Most of these 20 stars have only two observations (though some have three or four), so it is impossible to say which of them would fall in this range naturally and which would have been moved into this bin from binary motion.

\begin{figure}
\epsscale{1.2}
\plotone{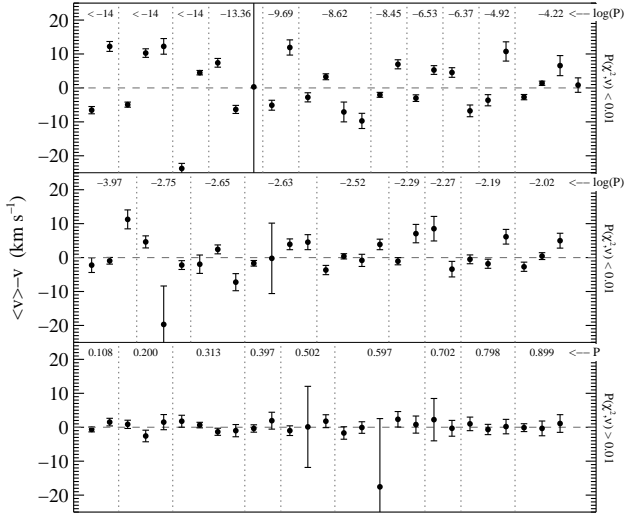}
\caption{Mean velocity for a star minus the individual velocity measures of that star. Observations are evenly spaced along the x-axis, and vertical gray dashed lines separate the velocities from one star to another. The top two panels of the plot show the 20 stars with $P(\chi^{2},\nu)<0.01$, and the bottom panel shows 9 stars with $P(\chi^{2},\nu)>0.01$ for comparison. $\log P(\chi^{2},\nu)$ for each star is listed at the top of the upper two panels and $P(\chi^{2},\nu)$ is listed at the top of the last panel.}
\label{chisqu_lowp}
\end{figure}

The amplitude of the velocity variability for these stars is illustrated in the top two panels of Figure \ref{chisqu_lowp}. A sample of nine stars that do not fall into this category, and thus have small velocity variability, are shown in the bottom panel for reference. The figure is essentially a glorified table; the difference between the weighted mean velocity of a star and its individual velocity measurements is plotted along the y-axis, and the x-axis simply serves as a way to separate one observation from another. Stars are distinguished by vertical gray dashed lines. Along the top edge of each panel, we list the probability corresponding to each star. For the first two panels, we listed the logarithm of $P(\chi^{2},\nu)$ since some of these probabilities are very very small, but in the last panel it is simply $P(\chi^{2},\nu)$. The small number of observations per star limits our ability to constrain the binary properties or to draw velocity curves, hence our reason for not plotting time along the x-axis.

The number of stars in the bin $P(\chi^{2},\nu)<0.01$ can be used to derive the lower limit for the binary fraction. If all 20 of the stars are binaries then the fraction would be 0.10. Given the variation in the number of stars per bin in the histogram, (i.e., 0--5) it is also plausible that only 15 of them are binaries, which produces a binary fraction of 0.08. We adopt the smaller of these as the minimum binary fraction for Leo II.

\section{Methodology}\label{sec_methods}

The method we use to find the binary fraction is to first generate a series of radial velocity Monte Carlo simulations that have the same velocity uncertainties and temporal observations as our real data. Then we use Bayesian analysis to compare the simulations to the data and ultimately determine which binary fraction can best reproduce the observed velocities in Leo II.

In Section \ref{bop}, we describe the seven binary orbital parameters that contribute to the radial velocity component of binary motion. In Section \ref{method}, we list the steps in the Monte Carlo simulations and explain how we can use an observable---called $\beta$---to perform Bayesian analysis. Section \ref{likelihood} gives the details of the Bayesian analysis, and Section \ref{ppds} shows how we extract the binary fraction from the posterior probability distribution.

\subsection{Binary Orbital Parameters}\label{bop}

\begin{figure}
\epsscale{1.1}
\plotone{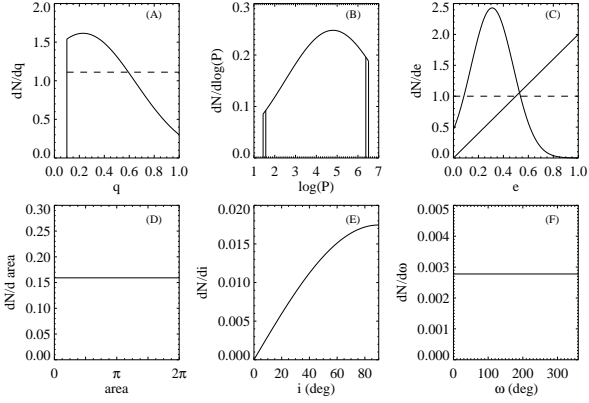}
\caption{Probability distribution functions for six of the binary parameters: mass ratio, period, eccentricity, area swept out since pericenter at time of first observation, inclination, and argument of periastron. Panels A and C show the distributions from \cite{duquennoy1991} as solid lines and the distributions from \cite{raghavan2010} as dashed lines. The eccentricity distribution in Panel C from \cite{duquennoy1991} depends on the period, so two functions are drawn.}
\label{bin_params}
\end{figure}

We start by writing the observed radial velocity associated with the orbital motion of a binary star, which can be expressed as
\begin{equation}
v_{r,orb} = \frac{q \sin i}{\sqrt{1-e^{2}}}\bigg(\frac{2\pi G m_{1}}{P(1+q)^{2}}\bigg)^{1/3}\big(\cos(\theta+\omega)+e\cos\omega\big)
\label{vorb}
\end{equation}
\citep[for a detailed derivation of this equation, see][Section 19]{green1985}. Note that this equation gives velocity relative to the system center of mass, which is what we observe. The seven parameters that characterize the orbital radial velocity are the mass of the primary ($m_{1}$), mass ratio ($q$), eccentricity ($e$), period ($P$), true anomaly ($\theta$), angle of inclination ($i$), and argument of periastron ($\omega$). Some are intrinsic to the system ($m_{1}$, $q$, $e$, $P$), and others depend on the observational circumstances ($\theta$, $i$, $\omega$). A diagram of the parameter distributions used in this analysis is shown in Figure \ref{bin_params}.

For the intrinsic parameters, we have adopted the distributions from \cite{duquennoy1991} and \cite{raghavan2010}, which are both based on Sun-like stars in the solar neighborhood. We have selected these distributions over the options in other papers \citep[e.g., ][]{fischer1992,reid1997,marks2011} so that we can perform a side by side comparison between our results and those of \citet{minor2013}. Furthermore, the distributions in these two papers for mass ratio and eccentricity are quite different, allowing use to get a sense of how big of a role they play in our analysis of the binary fraction. The lack of knowledge on the actual distributions for red giant stars in dSphs is the largest limiting factor in constraining the binary fraction in Leo II. Due to this shortcoming, additional distributions should be explored in subsequent analyses, especially those with different period distributions, as we will see in Section \ref{sec_leoiiresults}.

One exception to the distributions is $m_{1}$, which we fix at $m_{1}=0.8~M_{\odot}$. Our primary stars are all red giants and thus must have a mass around this value \citep{hargreaves1996}.

Next is the distribution of the mass ratio between binary stars, which is defined as $q=m_{2}/m_{1}$. The variable $m_{1}$ is the mass of the visible star and $m_{2}$ is the mass of the secondary star. We assumed the secondary star must be a non-remnant, non-giant star and must therefore have a mass $\le m_{1}$. It then follows that $q\le1$. We set the minimum mass ratio equal to $q_{min}=0.1$, such that the smallest companion is a hydrogen-burning object. The distribution for $q$ from \cite{duquennoy1991} takes the form
\begin{equation}
\frac{dN}{dq} \propto \exp(-\frac{(q-\mu_{q})^{2}}{2\sigma_{q}^{2}}),
\label{q_dist}
\end{equation}
where $\mu_{q}=0.23$ and $\sigma_{q}=0.42$. Alternatively, \cite{raghavan2010} finds a flat mass ratio distribution such that $\frac{dN}{dq} \propto const.$ Both of these distributions are plotted in panel A of Figure \ref{bin_params} and will be considered in this analysis.

We take the period distribution from \cite{duquennoy1991}, which has the log-normal form
\begin{equation}
\frac{dN}{d\log P} \propto \exp\bigg(-\frac{(\log P - \mu_{\log P})^{2}}{2\sigma_{\log P}^{2}}\bigg).
\end{equation}
For periods measured in days, $\mu_{\log P}=4.8$ and $\sigma_{\log P}=2.3$. \cite{raghavan2010} finds a similar distribution but with $\mu_{\log P}=5.03$ and $\sigma_{\log P}=2.28$. Since these two distributions are very similar, we choose to use the parameterization from \cite{duquennoy1991}. 
The minimum period possible for a binary corresponds to the minimum semi-major axis of the system, which is when the two stars are orbiting such that their surfaces are just out of contact. In our case, the primary is a red giant so the separation can be estimated as the radius of the larger star. Using a surface gravity of 10 cm s$^{-2}$ and a mass of 0.8 $M_{\odot}$ yields a radius of $a_{min}=0.21$ AU. When $q=0.1$ this corresponds to a period of $\log P_{min}=1.57$ (or 37.4 days), and when $q=1.0$ this is $\log P_{min}=1.44$  (or 27.8 days). These minima are plotted as the left two vertical lines in panel B of Figure \ref{bin_params}. For the maximum semi-major axis (and thus maximum period), we solve for the impact parameter of a star traveling through Leo II, such that $a_{max}=(\pi vtn)^{-1/2}$. $v=7.4$ \kms{} is the velocity dispersion \citepalias{spencer2017a} and $t=9\times10^{9}$ years is the average age of the main population of stars \citep{mighell1996}. Assuming an average star has mass 0.4 $M_{\odot}$ and $(L/L_{\odot})=(M/M_{\odot})^{4}$, then the average luminosity is 0.025 $L_{\odot}$. The central luminosity density of Leo II is $I_{0}=0.029$ L$_{\odot}$ pc$^{-3}$ \citep{mateo1998}, and so the volume that one star occupies is 0.88 pc$^{3}$. The number density is then $n=1.14$ stars pc$^{-3}$. \citep[For comparison, the number density of the solar neighborhood is about 0.13 stars pc$^{-3}$, ][]{chabrier2001}. This produces a maximum semi-major axis of  412 AU. Once again, when $q=0.1$ this corresponds to a period of $\log P_{min}=6.51$ log(days), and when $q=1.0$ this is $\log P_{min}=6.38$ log(days). These maxima are plotted as the right two vertical lines in panel B of Figure \ref{bin_params}.

The last intrinsic parameter is eccentricity, which has perhaps the least certain distribution of all. In principle this parameter can range from 0 to 1, but in practice the upper limit is often times smaller due to the constraints placed on period and mass ratio. The maximum eccentricity that keeps the stars from colliding is $e_{max}=1-(a_{min}/a)$, where $a$ is the semi-major axis that corresponds to $P$ and $q$ from above. \cite{duquennoy1991} found that the eccentricity distribution is a piecewise function that depends on period in such a way that 
\begin{equation}
 \frac{dN}{de} \propto \left\{ \begin{array}{cl}
\exp(-\frac{(e-\mu_{e})^{2}}{2\sigma_{e}^{2}}) & \textrm{if }1.08<\log P<3 \\
2e & \textrm{if } \log P>3 .
\end
{array}\right.
\label{e_dist}
\end{equation}
The shape of the first function was only based on 16 stars, so \cite{duquennoy1991} do not list parameter values. However, since we required a quantitative distribution, we took the mean $\mu_{e}=0.31$ and standard deviation $\sigma_{e}=0.17$ of these stars for the parameters of this distribution.
On the other hand, \cite{raghavan2010} claimed that the eccentricities for all stars with $\log P > 1.08$ followed a single flat distribution: $\frac{dN}{de}\propto const.$ Both studies agreed that the eccentricity for binaries with $\log P < 1.08$ would be 0 (circular) due to tidal interactions between the stars, but since we estimated the minimum period for red giants in Leo II to be $\log P_{min}=1.44$, we do not need to include this case in our analysis. 

The fifth parameter, $\theta$, is the angle between lines connecting the periastron to the focus and the focus to the star. This is called the true anomaly, and it is simply telling us the phase of the star within its orbit. Periastron is at $\theta=0\degree$ (or $360\degree$) and apastron is at $\theta=180\degree$. All other angles represent locations between these points and are dependent on the eccentricity. Due to its dependence on eccentricity, the probability density distribution for $\theta$ does not have an analytic solution. Instead, we pick the star's location within its orbit from the area swept out since periastron, and normalize it such that the area is 0 (or 2$\pi$) at periastron and $\pi$ at apastron. From Kepler's Second Law, we know that equal areas are swept out in equal times, and thus 
\begin{equation}
\frac{dN}{d~area} = const.
\label{area}
\end{equation}
Due to the way we have normalized it, this area is also known as the mean anomaly. We can then numerically solve for the true anomaly using the mean anomaly and the eccentricity. It is important to note that the mean anomaly for the first observation of the star can be drawn at random from Equation \ref{area}, but all subsequent mean anomalies that correspond to additional observations of a star are defined as $area=area_{1}+(2\pi\Delta t/P)$, where $\Delta t$ is the time elapsed since the first observation. 

The final two parameters concern the orientation of the system relative to our line of sight. The first of these is the angle of inclination, $i$, between our line of sight and the normal to the orbital plane. The probability distribution of the inclination angle is given by 
\begin{equation}
\frac{dN}{di} \propto \sin(i)
\end{equation}
where $i$ ranges from $0\degree$ (face on) to $90\degree$ (edge on). 

Last is the argument of periastron, which defines the angle of the ascending node of the orbit relative to the periastron point; this orientation is random and so $\omega$ takes on the simple form
\begin{equation}
\frac{dN}{d\omega} \propto const.
\label{omega_dist}
\end{equation}
where $\omega$ ranges from $0\degree$ to $360\degree$. Figure \ref{bin_params} plots the distributions for all six of these parameters. 

With ideal observing conditions (i.e. $area=0$, $i=90$, and $\omega=0$), a circular orbit ($e=0$), and a short period ($\log P_{min}=1.46$), the maximum change in velocity for a mass ratio of 1 and 0.1 is 81 \kms{} and 12 \kms{} respectively. For a long period ($\log P_{max}=6.39$) these values decrease to 1.8 \kms{} and 0.27 \kms{}. In practice, long-period binaries with these parameters will exhibit a change in velocity of around $10^{-4}$ \kms{} over a 19 year baseline.

\subsection{Method for Determining Binary Fraction}\label{method}

In the simulations that follow, we define the binary fraction, $f$, as the fraction of RGB stars that have a less massive (or equally massive) binary companion. The binary fraction ranges from 0 to 1.
Given the parameter distributions in Section \ref{bop}, the velocity measurement errors from the observations, and the Heliocentric Julian dates from the observations, model data were generated via Monte Carlo simulations as follows.
\begin{enumerate}
\item For a star in Leo II that has multiple observations, we selected it to be a binary or non-binary according to the binary fraction, $f$, under consideration.
\item If the star was determined to be a binary, we then selected a set of binary parameters from the distributions in Eqs. \ref{q_dist}-\ref{omega_dist}.
\item Then we calculated the orbital radial velocities of that star at all epochs when it was actually observed. These velocities were calculated from Eq. \ref{vorb} using the parameters chosen in Step 2. For a non-binary star, the orbital radial velocity was taken to be 0~\kms{}.
\item For both binary and non-binaries, gaussian deviates with standard deviation equal to the observational errors of the corresponding star and epoch were calculated and added to the  velocity of the star determined in Step 3. (In our analysis, we only cared about the change in velocity of the star over time, so we did not add additional radial velocity components from the motion of Leo II or the velocity dispersion since these are constant over the timescale of our observations.)
\item Steps 1-4 were repeated for all 196 stars in Leo II.
\item Steps 1-5 were repeated $\eta$ times to improve statistical certainty. For our case, we carried out $\eta=10,000$ trials per simulation.
\item Steps 1-6 were repeated for different binary fractions, from 0 to 1 in increments of 0.01.
\end{enumerate}

As a means of using our kinematic dataset of Leo II to determine the galaxy's binary fraction, we calculated the following statistic as a measure of the binary frequency of stars in the sample:
\begin{equation}
\beta = \frac{|v_{m}-v_{n}|}{\sqrt{\sigma^{2}_{m}+\sigma^{2}_{n}}}.
\end{equation}
In this relation, $v$ is velocity, $\sigma$ is the corresponding velocity error, and the subscripts indicate different observations for a single star\footnote{We also tried defining $\beta$ as $\frac{|v_m-<v>|}{\sqrt{\sigma^2_m+\sigma^2_{<v>}}}$, where $<v>$ is the average velocity of the star and $\sigma_{<v>}$ is the corresponding uncertainty. In one definition, we treated $<v>$ and $\sigma_{<v>}$ as the straight average and error; in a second definition, we considered them to be the weighted average and error. Both cases yielded similar results on the binary fraction. The first definition found a binary fraction that was different by only $\sim2\%$ while the second differed by $8\%$. These agree at the 0.5 $\sigma$ level. Furthermore, the width of the credible intervals differed by only 2\%--4\%.}. The number of $\beta$ calculations per star is equal to $n(n-1)/2$, where $n$ is the number of observations for that star. Since $n$ ranges from 2 to 7 in our sample, the number of $\beta$'s ranges from 2 to 21, and considering the distribution of $n$ in Figure \ref{nobs_deltat_year}, the total number of $\beta$'s is 723. When $\beta$ is computed from radial velocities in the observational data, we call it $\beta_{obs}$; when $\beta$ is computed from radial velocities in the model data, we call it $\beta_{mod}$.

A comparison between the distributions of $\beta_{obs}$ and $\beta_{mod}$ was then made using Bayesian analysis. The probability of Leo II having a binary fraction, $f$, given the data, $D$, and a set of models, $M$, is 
\begin{equation}
P(f|D,M) = \frac{P(D|f,M)P(f|M)}{P(D|M)} .
\label{bayesian}
\end{equation}
The variables $D$ and $M$ will be defined in the next subsection. The prior probability of the binary fraction in question, $P(f|M)$, is assumed to be uniform because there are no independent constraints on the binary fraction. The likelihood of the data given the models, $P(D|M)$, is a normalizing factor, which we selected such that the integral of the posterior is unity. Therefore, the posterior probability distribution, $P(f|D,M)$, is directly proportional to the likelihood, $P(D|f,M)$.

\subsection{Likelihood}\label{likelihood}

\begin{figure}
\epsscale{1.1}
\plotone{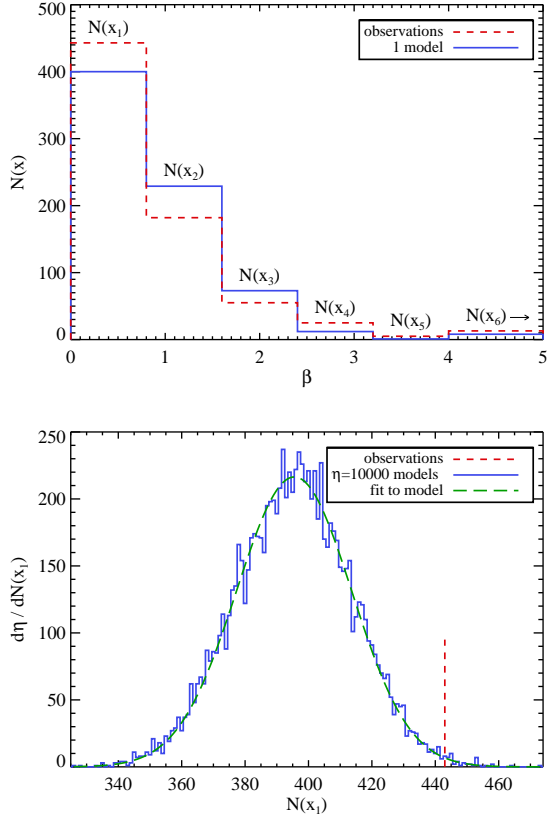}
\caption{The red dashed line is the histogram of $\beta_{obs}$ for the observations. For each Monte Carlo simulation, we generate histograms of $\beta_{mod}$ (top panel). For readability, we show only one of the $\eta=10,000$ simulations in the top panel as the blue solid line. The number of $\beta_{mod}$ that fall into bin one, $N(x_{1})$, for each of the $\eta$ simulations is then plotted in the bottom panel as the blue solid line. We mark the value in bin one for the observations as a vertical red dashed line. The models are fit using Equation \ref{skew}, which is shown as a green dashed line. This green line is then normalized and used as a probability density function to extract the probability of the observed galaxy being represented by this set of models for a given binary fraction. The process is repeated for all bins and all binary fractions to produce a PPD.}
\label{binary_example}
\end{figure}

Since calculating the likelihood is the most crucial part of the analysis, we include Figure \ref{binary_example} which illustrates two of the major steps in determining the likelihood and denotes key variables. In the top panel, we separated the $\beta$'s into six bins sorted by increasing $\beta$. The data $D$ is the number of $\beta_{obs}$ values in each bin $x$, and is shown as a red dashed line. For clarity and consistency, we redefine this as $N(x)_{obs}$. A similar histogram can be made for a set of $\beta_{mod}$ and is shown as a blue solid line. The number of $\beta_{mod}$ values in each bin $x$ is defined as $N(x)_{mod}$. We plotted only one histogram of $\beta_{mod}$ for readability, but there are actually $\eta$ in total since we performed $\eta$ Monte Carlo simulations for a given $f$, where $\eta$ was 10,000. The results in Figure \ref{binary_example} correspond to the distribution of $\beta$'s for the model in which the binary fraction, $f$, is 0.3 and the mass ratio and eccentricity distributions were constant.

The histogram bins are defined such that there are five between $0<\beta\le4$ with widths of 0.8, and a sixth bin that includes all $\beta > 4$. The $\beta=4$ limit reflects the fact that for $f=0$ (no binaries), we observed in our models very few instances where $\beta$ is this large. Placing this division at smaller $\beta$ would make it increasingly difficult to distinguish between $f=0$ and $f\ne0$. On the other hand, selecting a larger division would yield poorly populated bins since the vast majority of $\beta$'s are $<4$ even in cases with $f=1$, and would yield noisier results. Although much of the information on the binary fraction is contained within $\beta<4$, the number of $\beta$'s existing beyond this division is useful for ruling out (or confirming) small binary fractions because these cases would produce few large values of $\beta$. As a reference, less than $0.01\%$ of $\beta$'s exist in the last bin for the case of $f=0$ and there are less than $4\%$ $\beta$'s in this bin for $f=1$. For this reason, we collect all large $\beta$'s into a final bin to represent the tail of the $\beta$ distribution.

The bin width matters very little as long as the $\beta$ division is reasonably small, as in our case. Too large of a bin size will flatten the posterior, making it harder to distinguish $f$ from neighboring values of $f$, while too small of a bin size will produce a noisy posterior. We selected 0.8 because both of these effects were minimal for that value.

It should be noted that the bin width and division for the last bin can be changed somewhat before the aforementioned effects begin to take place. For example, we found that if we held the cutoff limit at 4, then we could drop the bin size down to 0.2 or increase it to 1.0 without seeing any statistically significant effects on the posterior. Alternatively, if we held the bin size at 0.8, we could change the cutoff value between 2.4 and 6.4 without it impacting the results.

In the bottom panel of Figure \ref{binary_example}, we have plotted $N(x_{1})_{obs}$ and $N(x_{1})_{mod}$. This is a single value for the observations so it is shown as a red dashed mark. For the models, there are $\eta$ values for this statistic (and $\eta=10,000$ in our case), so the resulting probability distribution function is plotted as a solid blue histogram. The probability density function that best fits $N(x_{1})_{mod}$ over all six bins takes the form of a skewed-normal distribution such that
\begin{multline}
\phi(N(x)|\mu,\sigma,\gamma) = \frac{1}{\sigma}\sqrt{\frac{2}{\pi}}\exp{\frac{-(N(x)-\mu)^{2}}{2\sigma^{2}}} \\ \times\int_{\infty}^{\gamma(N(x)-\mu)/\sigma} \exp{\frac{-z^{2}}{2}} \mathrm{d}z .
\label{skew}
\end{multline}
Here $\mu$ is the location, $\sigma$ is the scale, $\gamma$ is the skewness, and $z$ is a dummy variable. We performed a Levenberg--Marquardt least-squares fit of Equation \ref{skew} on $N(x_{1})_{mod}$, allowing all three parameters --- location $\mu$, scale $\sigma$, and skewness $\gamma$ --- to vary. The best fit with parameters $\mu_{mod}$, $\sigma_{mod}$, and $\gamma_{mod}$ is plotted in Figure \ref{binary_example} as a green long-dashed line and serves as the model $M$ for bin $x_{1}$. 

For a single bin $x_{1}$, the likelihood that the data $N(x_{1})_{obs}$ are given by the model $\phi(N(x_{1})_{mod}|\mu_{mod},\sigma_{mod},\gamma_{mod})$ and a binary fraction $f$ is $\phi(N(x_{1})_{obs}|\mu_{mod},\sigma_{mod},\gamma_{mod})$. The likelihood using all six bins is the product of the six individual likelihoods. Therefore, we can rewrite Equation \ref{bayesian} as 
\begin{equation}
P(f|D,M) \propto \prod_{x=x_{1}}^{x_{6}} \phi(N(x)_{obs}|\mu_{mod},\sigma_{mod},\gamma_{mod}).
\end{equation}
This is the the posterior probability for $f$, which can be repeated over all $f$ to find the posterior probability distribution (PPD).

\smallskip
\subsection{Characterizing the PPDs}\label{ppds}

To find the PPD statistic that best correlates with the binary fraction, we generated 11 sets of 200 mock galaxies with binary fractions between 0 and 1 in increments of 0.1, yielding 2200 galaxies in total. These galaxies have the same number of stars, number of observations per star, velocity errors, and observing cadences as our Leo II data. They are essentially just single Monte Carlo realizations and thus were generated using the same method as the simulations (see Steps 1-5 above). 

\begin{figure}
\epsscale{1.1}
\plotone{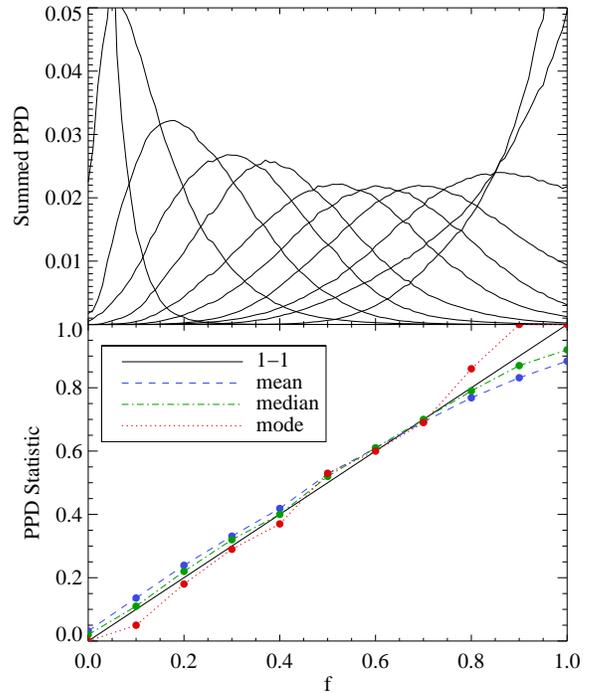}
\caption{Top: the PPDs of 200 mock galaxies with equal binary fractions were totaled to make eleven normalized curves. Bottom: the mean, median, and mode of the summed PPDs are plotted against the intrinsic binary fraction of the set of 200 mock galaxies. The mode (red dotted line) is biased toward lower binary fractions for $f\leq0.4$, whereas the mean (blue dashed line) is slightly biased toward higher binary fractions for $f\leq0.6$. The median (green dash--dotted line) is very slightly biased in the same direction as the mean, but with a magnitude less than $3\%$, it does the best job of reproducing the intrinsic binary fraction of the mock galaxies, and we therefore select as the statistical estimator for $f$.}
\label{statistics}
\end{figure}

We added up PPDs with the same assigned binary fraction to get a summed master PPD for each of the 11 binary fractions that we tested. These are shown in the top panel of Figure \ref{statistics}. The mean, median, and mode of these master PPDs are plotted against the true binary fractions in the bottom panel of Figure \ref{statistics}. The black line is the one-to-one line for which a statistic should follow if it perfectly matches the binary fraction that created it. The mean is the blue dashed line, the median is the green dash--dotted line, and the mode is the red dotted line. The mode underestimates the binary fraction for $f\le0.4$  but greatly overestimates it for $f\ge0.8$. The mean, on the other hand, overestimates the binary fraction for $f\le0.6$ and underestimates it for $f\ge0.8$. The median behaves in a way similar to the mean, but with a smaller bias of only a few percent. Therefore, we choose to use the median of the PPD as an indicator for the binary fraction of Leo II.

\section{Binary Fraction in Leo II}\label{sec_leoiiresults}

\begin{figure}
\epsscale{1.1}
\plotone{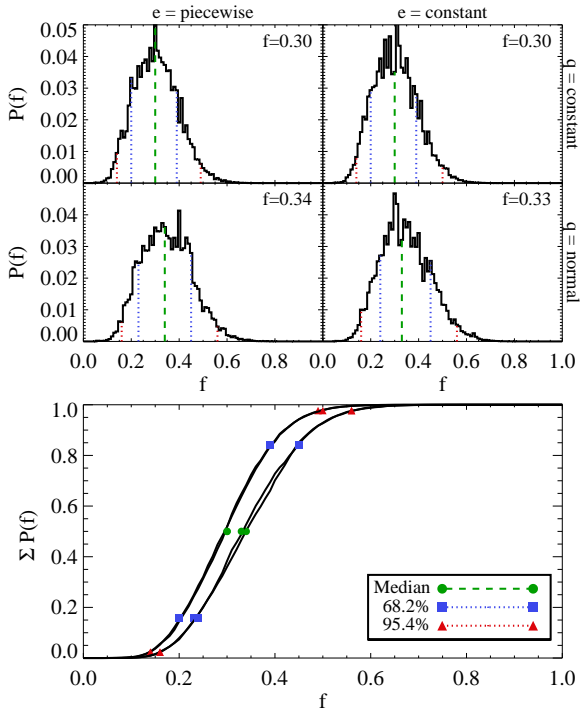}
\caption{Top: the posterior probability distributions of Leo II for four different combinations of the mass ratio and eccentricity distributions. The x-axis is binary fraction and y-axis is the probability that Leo II has that binary fraction. The median of the distribution is shown as a green dashed line, and is what we adopt as the binary fraction of Leo II. The 68\% credible interval is between the two vertical blue dotted lines and the 95\% credible interval is between the red dotted lines. These values are listed in Table \ref{table_ppds}. Bottom: the cumulative posterior probability distributions. The 68\% and 95\% credible intervals are repeated here and marked by blue squares and red triangles, respectively. The medians are the green circles.}
\label{bin_frac}
\end{figure}

Since we used two mass ratio distributions (normal in Equation \ref{q_dist} and constant) and two eccentricity distributions (piecewise in Equation \ref{e_dist} and constant), we have four different parameter combinations. The PPDs for these four sets of parameters of Leo II are shown in Figure \ref{bin_frac}. Median values range from 0.30 (constant $q$, constant $e$) to 0.34 (normal $q$, piecewise $e$). The medians are indicated with vertical green dashed lines in the PPDs in the top panel of Figure \ref{bin_frac} or a green dot in the bottom panel. The 68.2\% credible intervals are shown by blue dashed lines or blue squares, and the 95.4\% credible intervals are shown by red dashed lines or red triangles. Values for the medians and credible intervals are given in Table \ref{table_ppds}. Our highest estimate for the binary fraction of Leo II is $0.34^{+0.11}_{-0.11}$ for normal $q$ and piecewise $e$; the lowest estimate is $0.30^{+0.09}_{-0.10}$ for constant $q$ and constant $e$. Binary fractions above 0.63 or below 0.11 are strongly ruled out with $>99\%$ confidence regardless of the parameter distribution combinations.

Our use of multiple mass ratio and eccentricity distributions also allows us to determine two ways in which these affect the PPD. First, the PPD is very insensitive to the eccentricity distribution. The medians of posteriors that used a constant eccentricity distribution are larger than the medians of posteriors that used a piecewise eccentricity distribution by only 0.01. This result is illustrated by the fact that the cumulative PPDs group into nearly indistinguishable pairs in the bottom panel of Figure \ref{bin_frac}. Second, the mass ratio distribution plays a larger role in shaping the posterior than the eccentricity, though the effects are still minor. Posteriors that were built from a piecewise mass ratio distribution had medians that were 0.03--0.04 larger than posteriors with a constant mass ratio distribution. In intermediate stages of this analysis, we also considered a larger value for $\log P_{max}$ of 9.95 based on a static estimate, and found that this parameter could cause the median of the PPDs to increase by 0.10--0.12. We later discarded this value of $\log P_{max}$ as being unrealistically large, but would like to point out that the period distribution seems to have the biggest impact on the PPD. Our conclusions on the parameter distribution sensitivity are similar to those of \cite{minor2010} who found that the the posterior is also very sensitive to the position $\mu_{p}$ and width $\sigma_{p}$ values in the period function.

In the solar neighborhood, the binary fraction for main-sequence stars is estimated to range from around $2/3$ for F7-G9 type stars \citep{duquennoy1991} to $0.50\pm0.04$ for F6-G2 type stars and $0.41\pm0.03$ for G2-K3 type stars \citep{raghavan2010}. While some parameter distribution combinations provide better agreement than others, the values that we find for Leo II match the results from \cite{raghavan2010} within 1-2 $\sigma$. The agreement between Leo II and the solar neighborhood is not necessarily expected. One set of simulations predicted that dwarf galaxies should have larger binary fractions than the MW disk stars \citep{marks2011}. This motivates the need for better constraints on the binary parameter distributions for the most easily observable stars in dSphs (i.e. red giants).

\begin{deluxetable}{c c c c c}
\tablewidth{0pt}
\tablecaption{Median and Credible Intervals of PPDs\label{table_ppds}}
\tabletypesize{\scriptsize}
\tablehead{\colhead{$q$ Distribution}  & \colhead{$e$ Distribution} & \colhead{Median ($f$)} & \colhead{68.2\% Interval} & \colhead{95.4\% Interval}}
\startdata
normal   & piecewise & 0.34 & 0.23-0.45 & 0.16-0.56 \\
normal   & constant  & 0.33 & 0.24-0.45 & 0.16-0.56 \\
constant & piecewise & 0.30 & 0.20-0.39 & 0.14-0.49 \\
constant & constant  & 0.30 & 0.20-0.39 & 0.14-0.50 \\
\enddata
\end{deluxetable}

\cite{minor2013} reports the binary fraction in four MW dwarf spheroidals using data taken on Michigan/MIKE Fiber System at the Magellan/Clay telescope with $\sim1$ year baselines \citep{walker2009a}. Taking a similar but slightly different approach, they found the probability that each individual star was a binary and used likelihood analysis to extrapolate the overall binary fractions for the galaxies. Fornax, Sculptor, and Sextans all had similar fractions of $0.44^{+0.26}_{-0.12}$, $0.59^{+0.24}_{-0.16}$, and $0.69^{+0.19}_{-0.23}$, respectively, while Carina fell significantly below the others with a fraction of $0.14^{+0.28}_{-0.05}$ \citep{minor2013}. Given our highest and lowest estimates, Leo II seems to bridge the gap between the three galaxies with higher $f$ and the one with lower $f$. Other studies that comment on dwarf binary fractions discuss the fraction of stars that have velocity changes inconsistent with the velocity errors \citepalias[i.e.,][]{koch2007b}, or binary fractions over a shorter period range \citep[see, for example,][]{olszewski1996}. Because these are not global properties, we do not draw comparisons between them here.

Past kinematic studies of Leo II have concluded that the presence of binaries does not inflate the observed velocity dispersion by an appreciable amount \citepalias{vogt1995,koch2007b}. Although our binary fraction is larger than what was assumed in these studies, it does not change the conclusion: binaries cannot under typical circumstances artificially increase the true velocity dispersion of Leo II even when based on single-epoch kinematic measurements. We support this statement with a quick simulation using the equations and methods discussed in Sections \ref{bop} and \ref{method}. For a mock galaxy with an intrinsic velocity dispersion of 7.4 \kms{} and single-epoch observations for 196 stars (values equal to Leo II), the velocity dispersion will be increased by an average of 0.3 \kms{}. If the radial velocities are averaged over two epochs having a one year baseline, the dispersion is inflated by only 0.15 \kms{}. The length of the baseline does impact this effect, and an interval of 1-2 years has been found to be the optimal choice \citep{minor2010}. To further reduce the inflation, three or more epochs are required. Thus, in an era of multi-epoch observations with long baselines, it becomes safe to ignore the effects of binaries in Leo II and other similar dwarf spheroidals.

\subsection{Other Considerations: Heterogeneity}

\begin{figure}
\epsscale{1.1}
\plotone{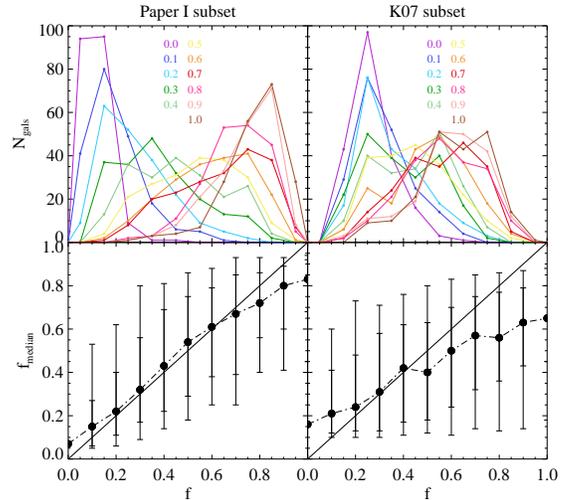}
\caption{Top: histograms of the binary fractions found for 200 mock galaxies with the actual binary fraction indicated by line color. For readability, they are drawn as points centered on bins with widths 0.1 connected by lines rather than traditional histogram stair-steps. Bottom: the median of each histogram is shown as a dot with 68\% and 95\% of all values falling within the small and large error bars respectively. If the analysis did a perfect job of recovering the binary fraction, then the dots would fall along the solid one-to-one line. Plots on the left were made using only data from \citetalias{spencer2017a} and plots on the right were made using only data from \citetalias{koch2007b}.}
\label{hetero}
\end{figure}

We also considered what effects, if any, the heterogeneity between the four datasets might have on the predicted binary fraction. Based on the case for constant eccentricity and mass ratio distributions, we assumed that the binary fraction for Leo II should always come out to be $0.30^{+0.10}_{-0.09}$ regardless of which subset of velocity data is used in the analysis. The binary fraction estimated with only data from \citetalias{spencer2017a} is $0.63^{+0.20}_{-0.22}$, whereas data from \citetalias{koch2007b} finds $0.13^{+0.29}_{-0.10}$. Both of these values are within the errors of the binary fraction found using the entire dataset, as is true for every other combination of data. Since the credible intervals on these numbers are so large, it could be possible for nearly any binary fraction to be consistent with 0.30. Therefore, we also ran Monte Carlo simulations to determine the probability that Leo II has a binary fraction of 0.30, but that the \citetalias{spencer2017a}/\citetalias{koch2007b} datasets individually predict it will be 0.63/0.13. The results of these simulations are shown in Figure \ref{hetero}, with \citetalias{spencer2017a} on the left and \citetalias{koch2007b} on the right. The top panels are composed of 11 histograms that each summarize the extrapolated binary fractions for 200 Monte Carlo galaxies with true binary fractions described by the color of the lines. The median of each of these histograms is plotted in the bottom panels; small and large error bars represent the ranges that 68\% and 95\% of the galaxies occupy respectively. If the binary fraction for these two subsets is in fact 0.30, then we should expect data with the same structure as \citetalias{spencer2017a} to recover binary fractions between 0.17 and 0.56 68\% of the time, or between 0.09 and 0.80 95\% of the time; for \citetalias{koch2007b} data we should expect binary fractions between 0.11 and 0.30 68\% of the time. The fact that \citetalias{spencer2017a} data and \citetalias{koch2007b} data found a wide range of binary fractions is thus expected. As long as the datasets are placed on the same velocity standard (as was done in Section \ref{sec_data}), there is no danger in mixing studies that have different velocity errors or different times elapsed between observations.

One additional feature in the bottom right panel is that the slope of the dashed line is shallower than the solid line, implying that some datasets, such as \citetalias{koch2007b}, can have a significant bias toward higher or lower binary fractions. In this case, the particular combination of small baselines, few repeat observations, and few stars lead to a set of $\beta$'s that did a poor job of constraining the binary fraction. Including other datasets will increase the number and range of $\beta$'s, thereby improving both the precision and accuracy of the binary fraction estimate. 

\subsection{Other Considerations: Velocity Errors}

This entire analysis has been completed using only three pieces of data: radial velocity, radial velocity uncertainty, and time of observation. There is very little error in the time of observation, and we have removed any errors in velocity to the best of our abilities by subtracting systematic offsets. However, it is more difficult to detect any errors in the velocity uncertainties.

To better understand how over- or under-reported velocity uncertainties could affect our results, we created two more Monte Carlo simulations with velocity errors either twice or half as large. In the first case, the PPD shows sharp spikes at a variety of binary fractions. The large errors mask any changes in velocity caused by binaries, making it impossible to tease out a binary fraction. Since we are seeing a much cleaner PPD for Leo II, we feel reassured that our velocity errors are not overestimated.

For the second case, the PPD does not have an unnatural shape. Instead, it yields a high probability that the binary fraction is 1. If any of the velocity uncertainties are underestimated it will push our binary fraction toward higher values. The change happens gradually. When the errors are only 10\% smaller, the binary fraction still works out to be $0.43^{+0.09}_{-0.10}$. This case is harder to rule out, however the distribution of $P(\chi^2,\nu)$ in Figure \ref{binary_chisqu} can help. When the errors are underestimated, the stars will cluster toward low $P(\chi^2,\nu)$. Alternatively, when the errors are overestimated, the stars will cluster toward high $P(\chi^2,\nu)$. Even in the case of 10\% smaller errors, stars begin to overpopulate the second to lowest bin. Our distribution is flat and shows no overpopulated bins (with the exception of the lowest bin being caused by binaries), so we are reasonably confident that the velocity uncertainties used in this paper are representative of the formal errors. Moving forward, we would like to emphasize the critical importance of robust error determination when exploring precision dynamics of dwarf galaxies.

\subsection{Consequences for Ultra-Faints}\label{sec_ultrafaints}

\begin{figure}
\epsscale{1.1}
\plotone{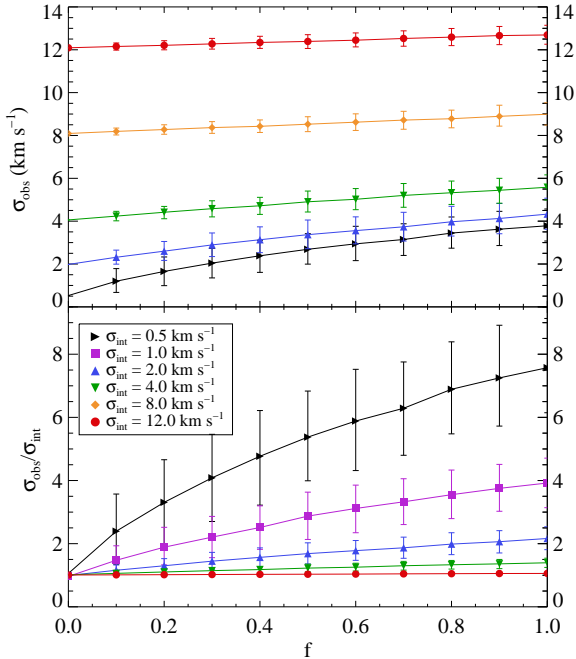}
\caption{Top: the observed velocity dispersion vs. binary fraction. Bottom: the ratio of observed to intrinsic velocity dispersion vs. binary fraction. Six mock galaxies are considered, each containing 100 stars with single-epoch observations and 1 \kms{} velocity uncertainties. The intrinsic dispersions are 0.5 \kms{} (black right-facing triangles), 1 \kms{} (purple squares), 2 \kms{} (blue upward triangles), 4 \kms{} (green downward triangles), 8 \kms{} (orange diamonds), and 12 \kms{} (red circles). As the binary fraction increases, so does the observed velocity dispersion. This effect is minimal for galaxies with high intrinsic dispersions, but for galaxies with low intrinsic dispersions, the observed dispersion can be 1.5--4 times that of the intrinsic dispersion for $f\sim0.3$. Models were generated using a normal mass ratio distribution and a piecewise eccentricity distribution.}
\label{binary_veldisp}
\end{figure}

As we have seen, binaries do not affect the velocity dispersion of Leo II and other classical dwarfs \citep{hargreaves1996,olszewski1996}. More recently, the problem has reemerged due to the possibility of binaries artificially inflating the dispersions of ultra-faint systems. In these cases, the dispersions appear to be $<4$ \kms{}, considerably smaller than in classical systems. To illustrate the severity of this issue, we completed yet another set of Monte Carlo simulations to explore the amplitude of this effect. We computed the observed velocity dispersion of six mock galaxies having intrinsic dispersions of 0.5, 1, 2, 4, 8, and 12 \kms{}. Each galaxy contained 100 stars with single-epoch observations and velocity measurement errors of 1 \kms{}. The results are shown in Figure \ref{binary_veldisp}. For galaxies with large intrinsic dispersions (red and orange lines in the figure), even binary fractions of 1 increase the observed dispersion by very little. On the other hand, galaxies with intrinsic dispersions of 0.5--2 \kms{} (black, purple, and blue lines) can have their observed dispersion inflated by a factor of 2--8 from a binary fraction of 1, or a factor of 1.5--4 for a more realistic binary fraction of 0.3. It is not likely that binaries are the sole contributors to the high velocity dispersions (and thus mass-to-light ratios) present in ultra-faints \citep{mcconnachie2010}, but even if the binary fractions are only $\sim0.3$, like what we find in Leo II, then they can play a non-negligible role in inflating the velocity dispersion. Furthermore, the stars typically observed in ultra-faints are subgiants or main-sequence stars rather than RGBs. These types of stars allow tighter binary orbits with shorter periods, which would increase the effects of binaries as well. 

\section{Summary and Conclusions} \label{sec_conclusion}

Multiple observations of several stars over many epochs made it possible to detect changes in radial velocities due to the presence of binaries. Coupling our data from \citetalias{spencer2017a} with \citetalias{vogt1995}, \citetalias{koch2007b}, and \citetalias{kirby2010} yields a total of 196 unique stars with two to seven observations between 1994 and 2013; the longest baseline for a single star was 19 years. 8\%--10\% of the stars have a small (e.g., less than 0.01) probability of exceeding their $\chi^{2}$, indicating that the change in velocity over time is significant and cannot be attributed to measurement errors. This corresponds to the detectable binary fraction in our sample. This value depends strongly on the baseline and number of observations per star and is therefore not a global property. 

To find the overall binary fraction for red giants in Leo II, we generated a suite of Monte Carlo simulations that sample from the seven binary parameters that define the orbital radial velocity for a binary star. We considered two mass ratio and eccentricity distributions, yielding four combinations of parameters. Using Bayesian analysis, we compared the simulations to the combined dataset and determined that the binary fraction for Leo II is around 0.30--0.34. The lowest recovered binary fraction was for a constant mass ratio distribution and a constant eccentricity distribution, which returned $f=0.30^{+0.09}_{-0.10}$ for a 68\% credible interval and $^{+0.20}_{-0.16}$ for a 95\% credible interval. The highest binary fraction was for a normal mass ratio distribution and piecewise eccentricity distribution, which returned $f=0.34^{+0.11}_{-0.11}$ for a 68\% credible interval and $^{+0.18}_{-0.22}$ for a 95\% credible interval. The results of all four simulations are listed in Table \ref{table_ppds}. Regardless of the parameter distributions, we can rule out binary fractions greater than 0.63 or less than 0.11 with 99\% confidence. Owing to the fact that the velocity dispersion of Leo II is large and our dataset is composed of stars with multiple observations, the effect of binaries on the velocity dispersion is negligible. 

While large systems like Leo II are little affected by binaries, these stars may play a bigger role in ultra-faints, particularly in cases of single or few observations. In our simulations, we found that dwarfs with low intrinsic velocity dispersions of 0.5--2 \kms{} could be observed to have dispersions 1.5--4 times larger than in actuality, given a binary fraction of 0.3. This effect further magnifies due to the extreme faintness of ultra-faints; the only way to increase kinematic samples in individual systems is to observe fainter stars, even down to the main sequence when feasible. In doing so, the period range and thus velocity amplitudes of binaries compared to larger red giant stars will increase. This has two important implications. First, it will be difficult to ever directly measure binary frequencies in ultra-faints. Second, the effects of binaries are necessarily amplified in ultra-faints not only because of their small dispersions, but also due to the increased impact binaries have on altering the velocities in the types of stars that need to be observed. 

Multi-epoch observations of ultra-faints are worth pursuing to directly explore their binary frequencies. Since it will be a while before the sample sizes of ultra-faints become large enough to accurately determine binary fractions on a case by case basis, an interim solution might be to correct the velocity dispersions using known binary fractions in brighter dSphs. The current results for dwarfs in the south use data that only span one year, but it will soon become possible to expand the analysis for Fornax, Sculptor, Sextans, and Carina as observations continue. Data for dwarfs in the north are already quite extensive so we plan to apply our method of determining the binary fraction to Draco and Ursa Minor. Combining our three galaxies with the four in \cite{minor2013} will give us a better picture of what the average binary fraction for dSphs is and if there are any dependencies on other galactic properties.
\medskip

\acknowledgements

The authors are very grateful to Josh Simon for allowing us to use his spectra to get velocities for the KG10 dataset, and to Jan Kleyna for contributing to the KK07 dataset. We thank Andy Szentgyorgyi and the Hectochelle team for their support over the past 12 years. We also thank the anonymous referee for helpful comments that improved this work. M.E.S. is supported by the National Science Foundation Graduate Research Fellowship under grant number DGE1256260. M.M. acknowledges support from NSF grant AST1312997. M.W. acknowledges support from NSF grants AST1313045 and AST1412999. E.O. acknowledges support from NSF grant AST1313006.

\end{document}